\newif\ifsubmode
\submodefalse
 
\ifsubmode
  \documentclass[12pt,preprint]{aastex}  
  \received{}
  \revised{}
  \accepted{}
 
\else
  \documentclass{emulateapj}  
  \slugcomment{Accepted to \apjl}
\fi

\newcommand{\kms}{km~s$^{-1}$}
\newcommand{\lsim}{\hbox{ \rlap{\raise 0.425ex\hbox{$<$}}\lower 0.65ex\hbox{$\sim$} }}
\newcommand{\gsim}{\hbox{ \rlap{\raise 0.425ex\hbox{$>$}}\lower 0.65ex\hbox{$\sim$} }}
\def \farcs{\hbox{$.\!\!^{\prime\prime}$}}
\def \farcm{\hbox{$.\!\!^{\prime}$}}

\shorttitle{Cl1604 Supercluster}
\shortauthors{Gal, Lubin \& Squires}

\begin{document}
\title{Pushing the Boundaries of the Cl 1604 Supercluster at $z\sim0.9$}

\author{Roy R. Gal\altaffilmark{1}, Lori M. Lubin\altaffilmark{2}}
\affil{Department of Physics, University
of California -- Davis, One Shields Avenue, Davis, CA 95616}
\author{Gordon K. Squires\altaffilmark{3}}
\affil{California Institute of Technology, M/S 314-6, 1200
E. California Blvd., Pasadena, CA 91125}
\altaffiltext{1}{gal@physics.ucdavis.edu}
\altaffiltext{2}{lmlubin@ucdavis.edu} 
\altaffiltext{3}{squires@ipac.caltech.edu}

\begin{abstract}

The Cl 1604 supercluster at $z\sim 0.9$ is known to contain at least
four distinct member clusters, separated in both projection and
redshift.  In this paper we present deep, multicolor wide-field
imaging of a region spanning $\sim45'$ on a side, corresponding
to $21~h^{-1}_{70}$ Mpc (physical) at the supercluster redshift.  We
select galaxies whose colors correspond to those of spectroscopically
confirmed cluster members in the $i'$ versus $r'-i'$ color-magnitude
diagram. Using an adaptive kernel, we generate a map of the projected
red galaxy density and identify numerous new candidate clusters that
are likely supercluster members. Assuming
that all of the density peaks are associated with the supercluster,
its transverse size is $\sim10~h^{-1}_{70}$ Mpc, which is still
significantly smaller than the nearly $93~h^{-1}_{70}$ Mpc depth in redshift space.

\end{abstract}


\section{Introduction}

The Cl 1604 supercluster at $z\sim0.9$ is one of the best studied
high-redshift large scale structures. It was initially detected as two
separate clusters in the plate-based survey of \citet{gho86}, with
redshifts and preliminary velocity dispersions measured by
\citet{pos98,pos01}. The proximity of the two clusters Cl 1604+4304
and Cl 1604+4321 in both radial velocity (4300~{\kms}) and position on
the sky ($17'$) prompted \citet{lub00} to perform deep multi-band
imaging with COSMIC \citep{kel98} on the Palomar 5-m telescope
covering the area between the two clusters. The imaging revealed
additional overdensities of red galaxies whose colors were consistent
with spectroscopically-confirmed, early-type galaxies in the two
original clusters \citep[see Figs. 2 and 4 of][]{lub00}, suggesting
that the clusters were part of a high-redshift supercluster. A total
of four distinct red galaxy density peaks were identified in a
contiguous area of $10\farcm4 \times 18\farcm2$. \citet{gal04} conducted a
large spectrographic survey using LRIS and DEIMOS \citep{fab03} on the Keck 10-m
telescopes to confirm the new cluster candidates and provide accurate
velocity dispersions for all four supercluster components. They
detected 230 total cluster members, confirming that all four
components were in fact supercluster members. These four clusters were
nevertheless well-separated both in projection and in redshift, and
have velocity dispersions ranging from 480 to 980 km s$^{-1}$,
corresponding to Abell richness classes of $R\sim0-2$.

The large radial depth of the supercluster ($\sim93~h^{-1}_{70}$ Mpc)
in comparison to the small area imaged with COSMIC
($4.8\times8.5~h^{-1}_{70}$ Mpc) suggested that this structure could
extend to a significantly larger area on the sky. To test this
hypothesis and search for new supercluster components, we undertook a
wide area, multi-band imaging survey using the Large Format Camera
\citep[LFC;][]{sim00} on the Palomar 5-m telescope.  In this paper,
 we present the results of this survey, including several new
candidate supercluster members.  We assume a $\Lambda$CDM cosmology
with $\Omega_m=0.3, \Lambda=0.7,$ and ${\rm H_0}=70~h_{70}~{\rm
km~s^{-1}~Mpc^{-1}}$.

\section{The Imaging Survey}

Two fields in the region of the Cl 1604 supercluster were imaged using
the LFC on the Palomar 5-m telescope. The LFC is a mosaic of six
$2048\times4096$ CCDs in a cross-shaped layout, mounted at prime
focus. The imaging area corresponds roughly to the unvignetted region
of a circle $24'$ in diameter. Data were taken on UT 2001 May 15 and
17 and UT 2004 April 26 and 27, with the camera installed in the
standard north-south orientation and all six CCDs in use. The imaging
data were taken in unbinned mode, with a pixel scale of $0\farcs182$
pixel$^{-1}$. Seeing on all nights was better than $1\farcs5$, with
average seeing of $1\farcs0$. Individual exposures of 450s were taken,
moving the telescope by up to $45''$ in each direction to alleviate
the image gaps due to the spaces between the CCDs. Data were taken at
two central pointings, using the Sloan Digital Sky Survey (SDSS)
$r'$,$i'$, and $z'$ filters. Only the night of UT 2004 April 27
was photometric, so individual exposures were taken in all three
filters at both pointings to allow calibration of the deep images. A
variety of standards from \citet{smi02} were also observed over a
range of airmasses on this night. Table~\ref{obslog} provides the
total integration times in each filter for the two pointings.
\setcounter{footnote}{0}

Data reduction was performed using the IRAF data reduction suite
\citep{tod86}, including the external package \textit{mscred}.\footnote{The Image Reduction and Analysis Facility (IRAF) is distributed by the National Optical Astronomy Observatory, which is operated by the Association of Universities for Research in Astronomy, Inc., under cooperative agreement with the National Science Foundation.}  We followed the
general mosaic imaging guidelines used by the NOAO Deep Wide Field
Survey \citep{jan04}, with modifications appropriate to our
data.\footnote{An extremely detailed data reduction description can be
found online at the Palomar Observatory instrumentation website, at
http://www.astro.caltech.edu/observatories/palomar/200inch/\\instruments.html}
The data were overscan-subtracted, bias corrected, flat fielded, and
fringe corrected in the standard way. Dark sky flats were generated by
combining the deep images; the application of this secondary flat
field results in images with less than $0.5\%$ residual sensitivity
variations across the field. Cosmic rays were detected using the
\textit{craverage} task, and a visual inspection of each image was
performed to identify any additional bad pixels or cosmic rays. All
bad pixels due to cosmic rays, instrument defects, saturation, bleed
trails, and satellite trails were masked. Astrometry was performed on
each image using the task \textit{msccmatch} to match detected
objects with the USNO-A2 catalog. Typically between 100 and 300
USNO-A2 stars are matched in each field and fitted with fourth-order
polynomials in both dimensions, resulting in rms errors of
$\sim0\farcs3$ in each coordinate. Individual exposures were then
projected onto the tangent plane using \textit{mscimage}, and all
exposures of a single pointing in a single filter were combined with
\textit{mscstack}. Additionally, all exposures of each field (in
$r'$, $i'$, and $z'$) were combined with \textit{mscstack} to produce an
extremely deep detection image. Images of photometric standards were
subjected to the identical reduction as target exposures.

\ifsubmode
\else
\begin{deluxetable}{crrccc}
\tablecolumns{6}
\tablewidth{0pc}
\tablecaption{LFC Imaging Log}
\tablehead{
\colhead{} & \multicolumn{2}{c}{Coordinates, J2000.0} & \multicolumn{3}{c}{Total Integration (sec.)} \\
\colhead{Pointing} & \colhead{RA}  & \colhead{Dec} & \colhead{$r'$} & \colhead{$i'$} & \colhead{$z'$}}
\startdata
1 & 16:03:53.5 & +43:21:33.3 & 3600 & 5850 & 3600 \\ 
2 & 16:05:05.7 & +43:04:18.4 & 4950 & 5850 & 3600 \\
\enddata
\label{obslog}
\end{deluxetable}
\fi

Photometric calibrations were derived from standard star magnitudes
measured in a variety of circular apertures using SExtractor
\citep{ber96}.  We solved for zero points, as well as color and
airmass terms using:
\begin{equation}
m_{true} = m_{inst} + A + Bsec(z) + C\times(color)
\end{equation}
where $A$ is the zeropoint, $B$ is the airmass coefficient, $sec(z)$
is the airmass at which the standard was observed, $C$ is the color
coefficient, and the color used is $r'-i'$ for the $r'$ and $i'$
data, and $i'-z'$ for the $z'$ data.

Object detection in the target frames was performed by running
SExtractor v2.3 in dual-image mode, using the ultra-deep image for
detection, while measurements were performed on the single-band
images. Photometry apertures are determined from the deep image, and
these same apertures are used for the individual filter images. We use
variable diameter elliptical apertures with major axis radius $2r_k$,
where $r_k$ is the Kron radius \citep{kro80,ber96}; the magnitudes
measured in these apertures are output as MAG\_AUTO in
SExtractor. This procedure results in catalogs in which magnitudes are
measured using identical apertures in all three filters, which
improves the measurement of galaxy colors by using the same physical
size for each galaxy in all filters \citep{lub00}. To calibrate
photometrically the deep exposures, shallower images taken on
photometric nights were SExtractor-ed, and matched to the deeper
data. Zero point offsets (typically $\sim0.1$ mag) were
derived and applied to the deep catalogs.  To remove spurious objects
around bright stars or near detector edges, as well as those with
uncertain photometry, we require detection in both the $r'$ and $i'$
images, and exclude objects with large photometric errors (MAG\_ERR$ >
0.2$ in either band), abnormally large semi-major axis (A $> 10$;
typically faint objects deblended from a very bright companion), and
zero isophotal area (objects in masked areas).

Because the two LFC pointings overlap slightly, the catalogs were
combined and objects detected in both pointings had one detection
removed. Additionally, there is a significant gap in the corner where
the two LFC pointings meet. To fill this area, we used existing data
from the COSMIC imager, taken as part of the survey presented in
\citet{lub00}. Figure~\ref{layout} shows the locations of the LFC and
COSMIC pointings.  The COSMIC images were taken using the Cousins $R$
and Gunn $i$ filters; to transform these data to the SDSS system, we matched
objects in the overlap region between the COSMIC and LFC fields, fitting
relations for both the $r'$ and $i'$ filters of the form

\begin{equation}
r' = A_rR + B_r(R-i) + C_r 
\end{equation}

The transformed COSMIC magnitudes show a $1 \sigma$ scatter of 0.1 mag
relative to the LFC magnitudes. The catalogs from the two LFC
pointings and the two COSMIC pointings were combined into a single
master catalog, deleting duplicate detections in overlap regions with
a preference for the slightly deeper LFC data. The final combined
catalog contains $\sim25,000$ unique objects.

\ifsubmode
\else
\begin{figure}
\plotone{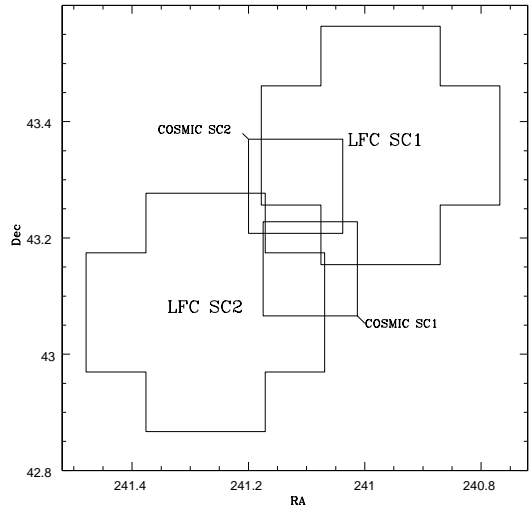}
\caption{Positions of the two LFC and two COSMIC pointings in the Cl 1604 region.}
\label{layout}
\end{figure}
\fi

\section{Cluster Detection}

The observed colors of early-type galaxies are strongly redshift
dependent. Even at $z\sim1.0$, the red sequence \citep{vis77} of
cluster ellipticals can be detected \citep{lub00,gla00}. Thus, searching for
overdensities of galaxies with the appropriate colors provides a
simple yet effective means of detecting clusters even at moderate to
high redshifts. In addition, we have obtained a substantial number of
redshifts of supercluster members \citep{gal04} that can be used to verify
the cluster galaxy colors.

Figure~\ref{cmd} plots the $r'$ versus $i'$ color-magnitude diagram
(CMD) of all objects in our imaging area. Spectroscopically confirmed
cluster members are superposed as large filled circles. The vertical
solid line shows the magnitude limit corresponding to the survey depth
of $r'=24.5, i'=24.5$. The horizontal and vertical dashed lines show
the color and magnitude cuts used to create the density maps in Figure
3. The right panel shows the color distribution of all objects with
$i'<24.5$ (dotted histogram) and the spectroscopically confirmed
supercluster members (solid histogram), scaled to the same peak
counts. The red sequence of supercluster members is clearly visible at
$r'-i' \sim 1.1$. Numerous bluer cluster galaxies are also evident;
these are likely galaxies with ongoing star formation and therefore
show strong [O \textsc{II}] $\lambda 3727$ emission redshifted into
the $r'$ filter. This hypothesis is consistent with the large fraction
($\sim85\%$) of spectroscopic members which show [O \textsc{II}]
emission in their spectra.  To perform cluster detection, we use only
those objects with $1.0\le r'-i' \le1.4$ and $20.5\le i' \le23.5$. The
color range was chosen to maximize the contrast of possible
supercluster members relative to fore- and background galaxies, as
seen in the right panel of Figure ~\ref{cmd}. The $i'$ magnitude limit
was selected so that both the COSMIC and LFC imaging catalogs are
reasonably complete in both $r'$ and $i'$. These cuts eliminate the
vast majority of foreground galaxies; only $\sim 1500$ objects satisfy
these criteria.

\ifsubmode 
\else
\begin{figure}
\plotone{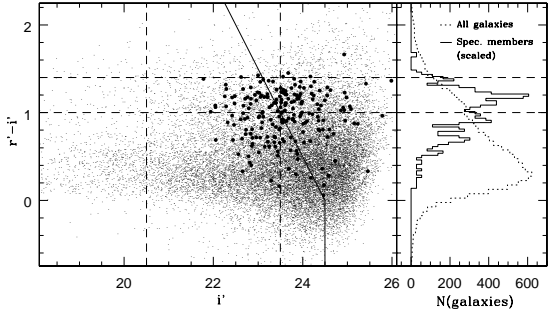}
\caption{The $i'$ versus $r'-i'$ color-magnitude diagram of all objects
in our imaging area. Spectroscopically confirmed cluster members are
overplotted as large filled circles. The vertical solid line shows the
magnitude limit corresponding to the survey depth of $r'=24.5,
i'=24.5$. The horizontal and vertical dashed lines show the color and
magnitude cuts used to create the density maps in Figure~\ref{akmap}. The right
panel shows the color distribution of all objects (dotted histogram) and spectroscopic members (solid histogram) scaled to the same peak number. }
\label{cmd} 
\end{figure}
\fi

\ifsubmode
\else
\begin{figure*}
\plotone{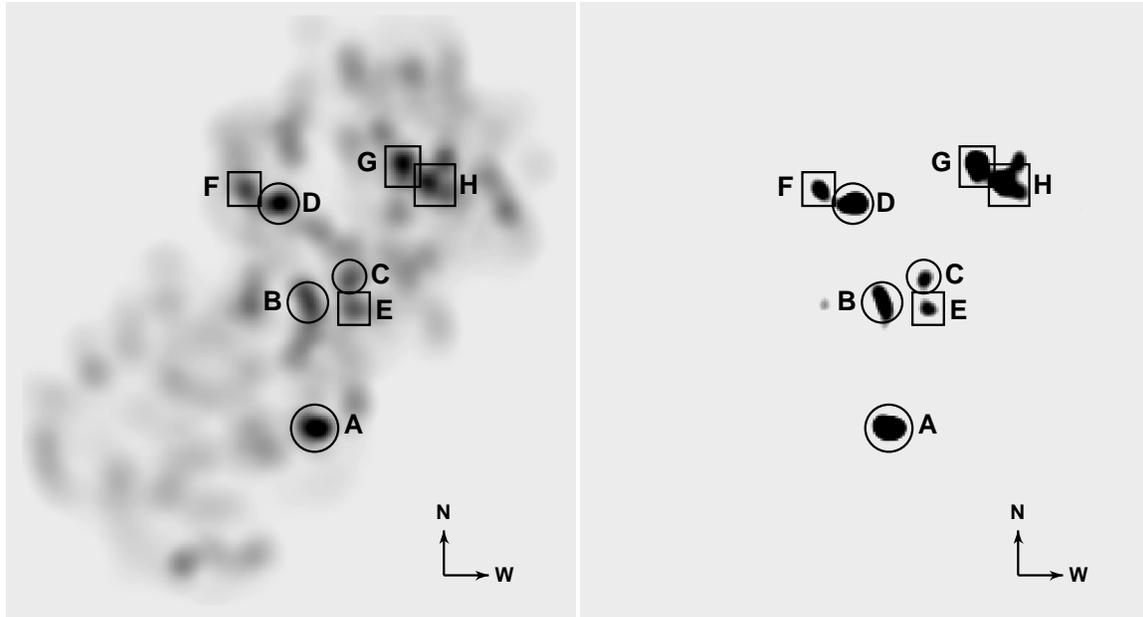}
\caption{Adaptive kernel density maps of galaxies meeting our color criterion 
in the Cl 1604 region. Previously
known clusters are marked with circles. New candidates are marked with
rectangles. The left panel shows the overall structure of the
region. The right panel shows only areas with $\ge14000$ red galaxies
deg$^{-2}$.  Each map covers a $39\farcm8 \times 43\farcm6$ area.}
\label{akmap} 
\end{figure*}
\fi

Using the color-selected galaxies, we produce an adaptive kernel
surface density map \citep{sil86,gal03}, with $10''$ pixels and an
initial smoothing window of $128''$, corresponding to 1
$h_{70}^{-1}$ Mpc at the supercluster redshift. The adaptive kernel
uses a two-stage process to generate a density map. It first produces
an initial estimate of the galaxy density at each point in the map
using a fixed grid. It then uses this pilot estimate to apply a
smoothing kernel whose size changes as a function of local density,
with a smaller kernel at higher density. The smoothing size is chosen
to separate the individual clusters in the supercluster without
resolving much of the substructure in the clusters. The left panel of
Figure~\ref{akmap} shows the density map for the color-selected
objects from the combined LFC and COSMIC data, covering an area of
$39\farcm8 \times 43\farcm6$. The four previously detected clusters at
$z\sim0.9$ are marked with circles and labeled with the same lettering
scheme as \citet{gal04}. These four clusters are clearly detected as
significant density peaks. However, there are a number of additional
density enhancements of similar significance to the known members,
especially to the northwest of the originally imaged region. To
confirm that these structures are not spurious, we generated several
density maps using a variety of smoothing kernels corresponding to
$0.5-1.5~h_{70}^{-1}$ Mpc, as well as somewhat narrower and broader
color cuts. Similar structures were visible in all such maps. We note
the existence of numerous less significant density peaks and filaments
in this map. These may be filamentary structure and possible
very-low-mass systems associated with the supercluster.

The right panel of Figure~\ref{akmap} shows the same density map, but
only pixels with red galaxy surface densities of $\ge$14,000 galaxies
deg$^{-2}$ are shown. This threshold is set to include all of the
spectroscopically confirmed clusters. At least four additional
structures, labeled ``E,'' ``F,'' ``G'', and ``H'', are above
this threshold. We note that this choice of threshold is subjective;
as mentioned before, many other overdensities are apparent. However, the
spectroscopic observations indicate that the poorest of the four
confirmed clusters has a velocity dispersion of only 489 km s$^{-1}$,
typical of low mass clusters \citep{bah03}. Therefore, the chosen
density contrast will select galaxy concentrations consistent with
even rather poor clusters. We also ran SExtractor on the density map
with detection parameters fine-tuned to detect the four known
clusters. This automated detection procedure picks up the same density
peaks as those found visually.

It is unclear from this map whether E and F are unique
clusters, or substructures related to the known members C and
D, respectively.  Unfortunately, the newly detected structures
are outside the region for which we have obtained spectroscopy, so no
conclusions about their independence can be drawn at this
time. Similarly, G and H are very close in projection and
may be components of a merging system. They are, however, clearly separated
from the other supercluster members. Their nearest neighbor is over 4
$h_{70}^{-1}$ Mpc distant. Future DEIMOS spectroscopic observations
are planned to determine supercluster membership for the newly
discovered structures.

\section{Cluster Properties}

From the multicolor imaging data we are able to examine a limited
number of cluster properties for each candidate. The richness is
estimated by counting the number of red galaxies
($1.0\le(r'-i')\le1.4$) in the magnitude range where both the COSMIC
and LFC data are complete ($20.5\le i'\le23.5$).  If all of the new
candidate clusters are at the same redshift as the known ones, then
this method samples the same (albeit bright) portion of the luminosity
function and gives the relative richnesses of the clusters, subject to
variations in the galaxy colors among the clusters. The computed
richnesses, along with the cluster coordinates, are provided in
Table~\ref{clusprops}. These coordinates are measured from the peaks
in the red galaxy density map, and are slightly offset from the
spectroscopic centers reported in \citet{gal04}. The photometric and
spectroscopic centers of clusters A, C, and D agree to within $1''$;
the photometric center of cluster B is $\sim30''$ (0.25 $h_{70}^{-1}$
Mpc) south of the spectroscopic center, likely because this cluster
appears significantly elongated to the south. The redshifts and
velocity dispersions from
\citet{gal04} for the previously detected supercluster
 members are also given. We note that cluster C has comparable richness to three other previously known clusters, despite having a much lower velocity dispersion. This cluster was only partially imaged in the earlier survey and thus may not have complete spectroscopic coverage. 

\ifsubmode
\else
\begin{deluxetable*}{crrrrr}
\tablecolumns{6}
\tablewidth{0pc}
\tablecaption{Cluster Properties}
\tablehead{
\colhead{Cluster} & \colhead{RA}  & \colhead{Dec} & \colhead{Richness} &\colhead{$\sigma_{los}$ {\tiny(km s$^{-1}$)}} &\colhead{Redshift}}
\startdata
A & 16:04:20.9 & +43:04:40.6 & 47 & 962 & 0.9001 \\ 
B & 16:04:23.6 & +43:14:30.5 & 38 & 719 & 0.8652 \\
C & 16:04:06.5 & +43:15:40.4 & 39 & 489 & 0.9350 \\
D & 16:04:35.7 & +43:21:19.6 & 42 & 640 & 0.9212 \\
E & 16:04:05.0 & +43:13:31.9 & 38 & ... & ... \\
F & 16:04:49.2 & +43:22:19.5 & 37 & ... & ... \\
G & 16:03:45.4 & +43:24:07.0 & 53 & ... & ... \\
H & 16:03:28.5 & +43:24:16.5 & 58 & ... & ... \\
\enddata
\label{clusprops}
\end{deluxetable*}
\vskip 0.3in
\fi

We also construct CMDs for each cluster, using a circular region with
radius 1 $h_{70}^{-1}$ Mpc centered on each density peak. These are
shown in Figure~\ref{cluscmds},along with the combined CMD for all
eight clusters. To the right of each CMD, we show the color
distribution in each field as the solid histogram; this is simply the
result of collapsing each CMD along the magnitude axis. For
comparison, the dotted histogram shows the color distribution in a
relatively blank region of the imaging data, scaled to an area with radius of 1 $h_{70}^{-1}$ Mpc. The enhancement of red
cluster galaxies is clearly visible in these individual CMDs and color
distributions.

\ifsubmode
\else
\begin{figure*}
\plotone{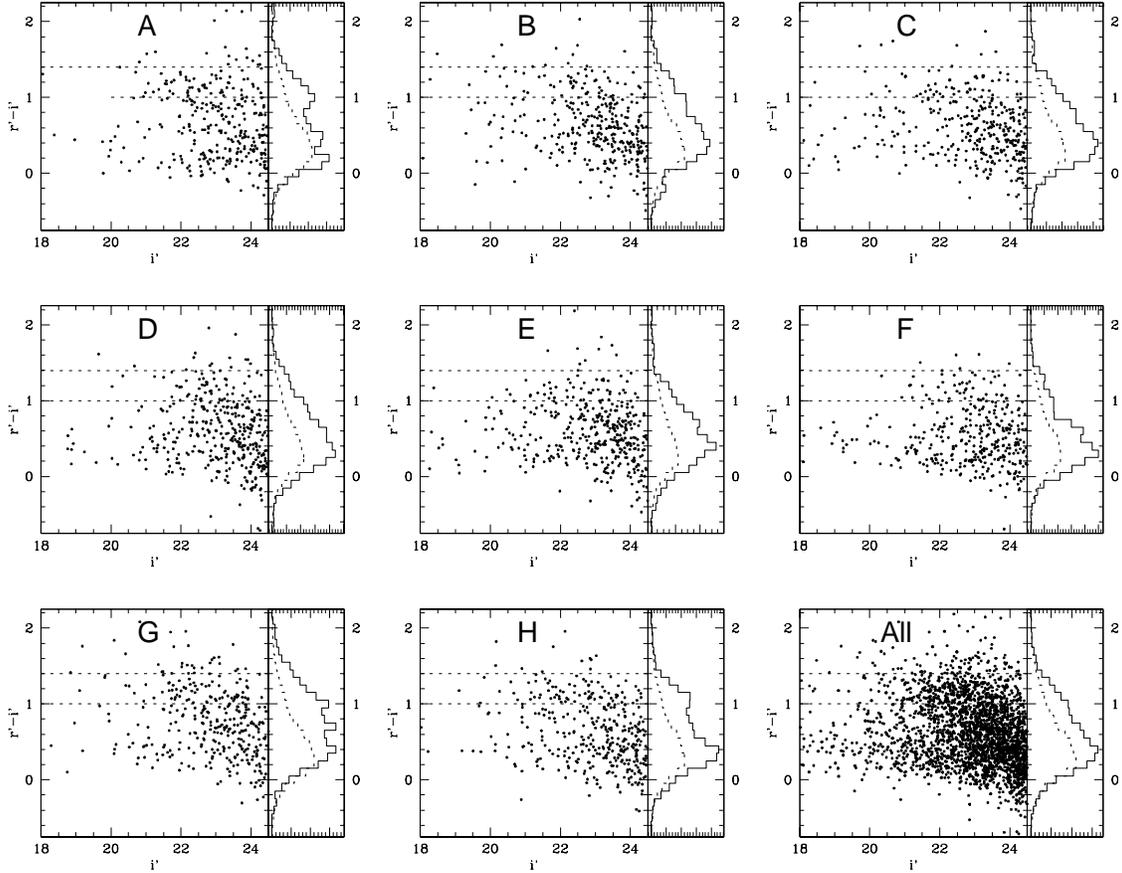}
\caption{The $i'$ versus $r'-i'$ color-magnitude diagrams within a 1 $h_{70}^{-1}$ Mpc radius for the four previously known clusters 
and four new candidates, as well as the eight cluster areas 
combined. The dashed horizontal lines indicate the color range used for 
cluster detection. Histograms of $r'-i'$ color for each cluster are plotted as the solid histogram to the right of each CMD, with the dotted histogram representing the color distribution in a blank sky region, scaled to the same area as the cluster
histogram.}
\label{cluscmds} 
\end{figure*}
\fi

\section{Discussion}

If the red galaxy overdensities detected in this imaging survey are
truly members of the Cl 1604 supercluster at $z\sim0.9$, we have
identified by far the largest known structure at redshifts approaching
unity. The largest projected separation (AG) is approximately
$9.6~h^{-1}_{70}$ Mpc, still much smaller than the apparent
$93~h^{-1}_{70}$ Mpc depth. It appears that we have detected the
southeast limit of the supercluster; no significant overdensities are
visible within $\sim10~h^{-1}_{70}$ Mpc directly to the east of A. We
cannot rule out the structure continuing to the north/northeast or
southwest. Most of the highest density peaks, including the four known
clusters, are surrounded by smaller, lower density enhancements,
suggesting a filamentary nature that is typical of large scale
structure in simulations \citep{evr02}. The large range of densities
in this supercluster makes it an ideal laboratory to study the
dependence of galaxy properties and their evolution on their physical
environment.

We note that the apparent 8:1 axial ratio of this supercluster is
comparable to or even less than that found in
simulations. \citet{sha04} provide an excellent discussion of shape
measurements for filaments and voids in numerical simulations with
$\Lambda$CDM cosmologies, demonstrating that superclusters typically
have non-trivial, highly filamentary topologies with
order-of-magnitude or greater differences in the lengths of their
dimensions (see their Figures 14 and 15). Furthermore, any complex
filament will be easier to detect if it is viewed in projection along
its longest axis. If the angular separation between the sufficiently dense
regions of a supercluster becomes larger than the contiguously imaged
area in a survey, then it will only be detected as a single cluster.

Our results demonstrate the efficiency of deep, moderately large area
imaging in the vicinity of known clusters for detecting further
structures. Hubble volume simulations show that similar superclusters
are not uncommon at $z\sim1$ \citep{evr02}. Therefore, we have
undertaken a wide-field imaging survey of a large sample of $z > 0.5$
clusters to search for other large scale structures at these
cosmologically significant redshifts. Using color cuts corresponding
to early-type galaxies at a variety of redshifts we can also detect
clusters at distances other than that of the target, in a manner
similar to the cut-and-enhance technique of \citet{got02} and the Red
Sequence Cluster Survey \citep{gla00}.

\acknowledgements

We thank M. Hunt for early assistance with LFC data reduction.We also
thank the anonymous referee for useful comments that clarified the
text. This research was supported in part by grant HST-GO-08560.05-A
from the Space Telescope Science Institute. This work is based on
observations taken with the Hale Telescope at Palomar Observatory as
part of a continuing collaboration the California Institute of
Technology, NASA/JPL, and Cornell University.

\end{document}